\documentclass{kapproc} % Computer Modern font calls
\usepackage{xspace}
\usepackage{defs}
\let\footnote\savefootnote
\let\footnotetext\savefootnotetext 
\let\footnoterule\savefootnoterule 
\kluwerbib

\RequirePackage{graphicx}%
\RequirePackage{epsfig}%

\def\eg{{\it e.g.}}
\def\ltsima{$\; \buildrel < \over \sim \;$}
\def\simlt{\lower.5ex\hbox{\ltsima}}
\def\gtsima{$\; \buildrel > \over \sim \;$}
\def\simgt{\lower.5ex\hbox{\gtsima}}
\def\fesc{{$\langle f_{esc}\rangle$}\xspace}

\def\H2{H$_2$\xspace}

\def\nor{``large-halo''\xspace}

\def\p3{``small-halo''\xspace}

\def\Mpc{$h^{-1}$~Mpc\xspace}

\begin{document}

%\articletitle{DSphs/first-galaxies connection}
\articletitle{Dwarf Sphs/First-galaxies\\ connection}
\author{Massimo Ricotti}

\affil{
CASA, University of Colorado, Boulder CO 80309 and\\
IoA, University of Cambridge, UK, CB3 0HA}
\email{ricotti@ast.cam.ac.uk}

%% optional, to supply a shorter version of the title for the running head:
\chaptitlerunninghead{Dwarf Sphs/First-galaxies connection}

\begin{abstract}
  I analyze the properties of the first galaxies in cosmological
  simulations with radiative feedback. Preliminary results indicate
  similarities with the observed properties of the bulk of dwarf
  spheroidal galaxies (dSphs) in the Local Group and Andromeda. I
  briefly discuss observational tests that could help in
  understanding the impact of a population of small primordial objects on
  the cosmic evolution.
\end{abstract}

\section{Introduction \label{sec:int}}

Ricotti, Gnedin and Shull 2002a,b \nocite{RicottiGSa:02}
\nocite{RicottiGSb:02} have studied radiative feedback processes that
regulate the formation of the first galaxies. Contrary to normal
galaxies, the global star formation in these objects is self-regulated
on cosmological scales. This happens because internal and external
radiative feedback processes are important in triggering or
suppressing their ability for form stars. Typically star formation in
the first galaxies is bursting and the emitted ionizing photons remain
confined in the denser filaments of the intergalactic medium (IGM),
preventing a complete IGM reionization. The main parameter that
regulates the star formation is \fesc: the escape fraction of ionizing
photons. In this talk I present preliminary results on the properties
of the first galaxies in our simulations. I try to understand the
differences and similarities with observed dSphs and discuss the
observational consequences of the existence of such a population of
small primordial galaxies \footnote{By definition these
  ``microgalaxies'' form in dark matter (DM) ``minihalos'' with masses
  $M_{DM} \le 10^8$ M$_\odot$.  If the gas is of primordial
  composition (metal free), molecular hydrogen is the only coolant
  available to form dense gas clouds and the first stars
  (Population~III).  There is not consent on a unique name for this
  population of primordial objects, sometimes also called PopIII
  objects or ``small-halo'' galaxies.}.  DSphs have masses comparable
to those of the first galaxies, but this does not necessarily imply
that they are relics of primordial objects.  It is possible that part
or all of the observed dSphs are galaxies formed later from larger
Dark Matter (DM) halos, subsequently stripped of part of their DM
content.  Our goal is to distinguish between these two formation
scenarios comparing observed properties with simulated properties.
  
\section{First galaxies versus Dwarf Spheroidals}

\begin{figure}[htb]
%\epsscale{0.8}
\plotone{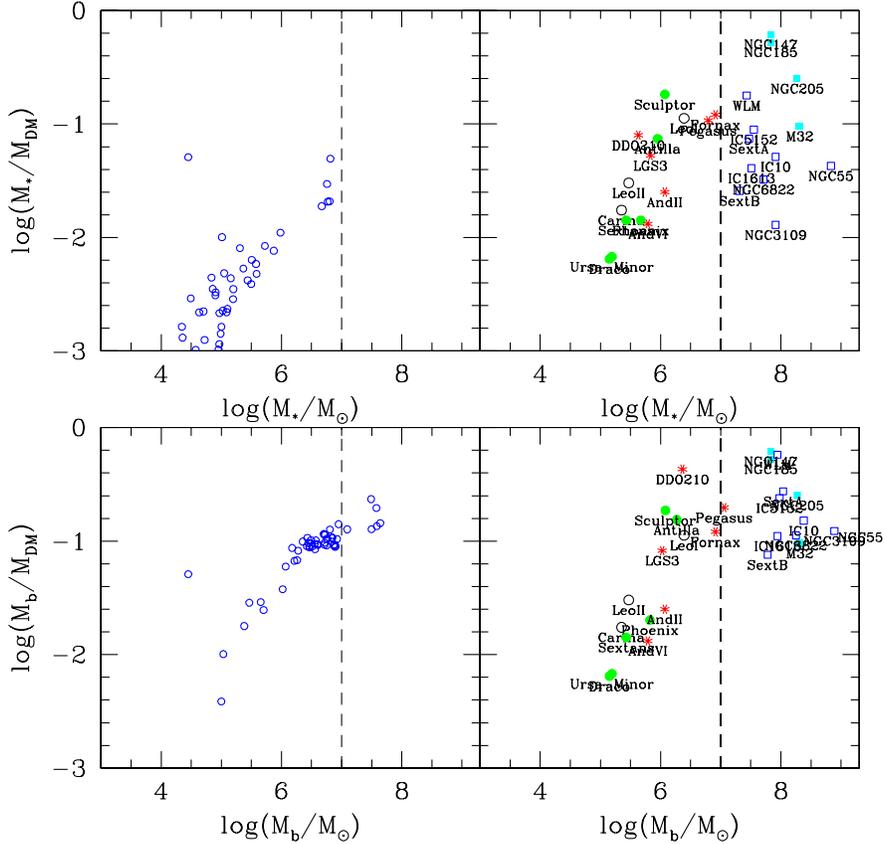}
\caption{\label{fig1} Comparison between properties of simulated
  primordial galaxies (left) at z=10 and dwarf galaxies (right) at
  z=0. (Top) Star fraction $f_*=M_*/M_{DM}$ as a function of $M_*$.
  (Bottom) Baryon fraction $f_b=M_b/M_{DM}$ as a function
  of $M_b=M_*+M_{gas}$. }
\end{figure}

\begin{figure}[htb]
%\epsscale{0.8}
\plotone{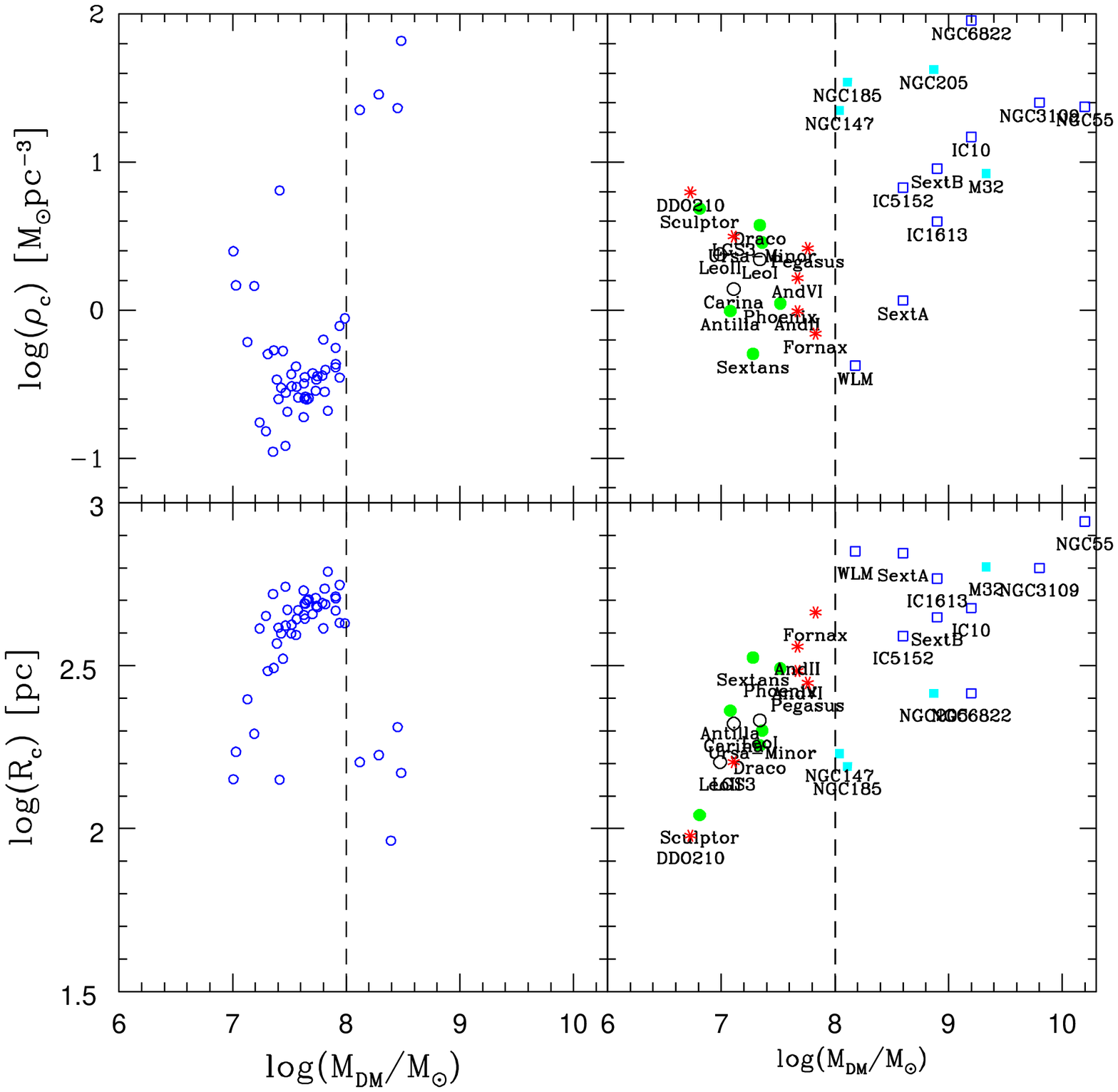}
\caption{\label{fig2} Same as in Fig.~\ref{fig1}. (Bottom) Core radius
  $R_c$ (pc) as a function of dynamical
  mass($M_{DM}=167\beta R_c\sigma_0^2$, where $\beta=8$ and $\sigma_0$
  is the central velocity dispersion in km/s). (Top) Core density
  ($\rho_c=M_{DM}R_c^{-3}$) as a function of dynamical mass.}
\end{figure}

Perhaps the most remarkable characteristic of dSphs is the large
velocity dispersion of their stars compared to their total visible
mass. This observation is often interpreted as the presence of a dark
halo that dominates dynamical motions down to the very center of the
galaxy. DSphs are either gas poor or do not contain gas at all
(NGC147), with a few exceptions (\eg, Sculptor perhaps has a gas cloud
in the other halo). The stars in dSphs are usually fitted with
exponential luminosity profiles or low concentration King profiles.
Most of the dSphs have experienced one or more bursts of star
formation about 15Gyr ago and do not present recent star formation
episodes.  Exceptions to this rule are Carina, LeoI and LeoII. Well
known properties of dSphs are the luminosity-metallicity relationship
and the luminosity-mass to light ratio relationship. When
spectroscopic metallicities of single stars are available, the spread
in metallicity of the stars is often very large (\eg, Draco
$-3<[Fe/H]<-1.5$).  Figs.~\ref{fig1}-\ref{fig2} compare some of the
aforementioned relationships with the results of one of our
simulations (labeled ``p3'' in Table~\ref{tab:2}) at redshift z=10.
The left panels show the simulated galaxies with $M_* > 10^4$
M$_\odot$. Only few normal galaxies with $M_{DM} > 10^8$ M$_\odot$ are
present in the simulation at z=10, but they appear to have different
properties. The right panels show both dSphs and dwarf Irregulars in
the Local Group and Andromeda. The symbols have the following meaning:
very old dSphs (filled circles); old dSphs (asterisks); dSphs with a
young stellar population (open circle); dwarf Irregulars (open
squares); dwarf Ellipticals (filled squares).  The mass to light ratio
of old dSphs should be $M_*/L_V \approx 1.2$ (typical for globular
clusters). Smaller values are expected for dwarf Irregulars and
dSphs with more recent star formation episodes. Here we adopt
$M_*/L_V=0.6$, because gives a maximum baryon fraction similar the
cosmic value (but the scatter is large). 
%Luminosity has been overestimated because of field stars contamination.
\def\tabone{
\begin{deluxetable}{lcccccc}
%\footnotesize
\tablecaption{dSphs properties.\label{tab:1}}
\tablewidth{0pt}
\tablehead{
\colhead{name} & \colhead{$M_{DM}$\tablenotemark{a}} & \colhead{$M_*$} &
\colhead{$R_c$} & \colhead{$R_t$} & \colhead{star ages} & \colhead{gas} \\
\colhead{} & \colhead{M$_\odot$} & \colhead{M$_\odot$} & \colhead{pc}
& \colhead{pc} & \colhead{Gyr} & \colhead{\%}
}
\startdata
 . & . & . & & & &
\enddata
\tablecomments{....}
\tablenotetext{a}{...}
\end{deluxetable}
}
%\tabone
\def\tabtwo{
\begin{deluxetable}{lcccccc}
%\footnotesize
\tablecaption{Simulation parameters.\label{tab:2}} 
%\tablewidth{0pt} 
\tablehead{\colhead{RUN} & \colhead{$N_{box}$} & \colhead{DM Mass Res.} &
\colhead{SED} & \colhead{$\epsilon_{UV}$} & \colhead{$\epsilon_*$} & \colhead{\fesc} \\
    \colhead{} & \colhead{} & \colhead{$h^{-1}$ M$_\odot$ } &
    \colhead{} & \colhead{$(10^{-5})$} & \colhead{} & \colhead{}}
\startdata
  p2-2 & 128 & $3.94\times 10^4$ & Pop. II  & 1.1 & 0.05 & 1\%\\
  p3   & 256 & $4.93\times 10^3$ & Pop. III & 2.5 & 0.1  & 10\% \\
\enddata 
\tablecomments{Parameter description: $N_{box}^3$ is the number of
    grid cells (the box size is 1 \Mpc). $\epsilon_*$ is the star
    formation efficiency, $\epsilon_{UV}$ is the ratio of energy
    density of the ionizing radiation field to the gas rest-mass
    energy density converted into stars (depends on the IMF), and
    \fesc is the escape fraction of ionizing photons from the
    resolution element.}
\end{deluxetable}
}
\tabtwo
% \begin{figure}[htb]
% %\epsscale{0.8}
% \plotone{CAT/dw_zm.eps}
% \caption{\label{fig3}Same as in Fig.~\ref{fig1}. (Top) Star
%   metallicity as a function of stellar mass $M_*$. (Bottom) Gas
%   metallicity as a function $M_*$.}
% \end{figure}
% \begin{itemize}
% \item[yes] DM dominated.
% \item[yes] gas poor [Sculptor (blob of gas in the halo) and NGC 147 (no gas)]
% \item[yes] mainly old stars (but some has distinct episodes of SF: Carina
%   has one quite recent)
% % MCs can store and release gas
% \item[yes] luminosity profile: low concentration King's profile or exponential.
% \item[yes] luminosity-metallicity relation, metallicity spread
%Draco $-3<[Fe/H]<-1.5$ average $[Fe/H]=-2\pm 0.15$\\
%In the local group:
%Luminosity/metallicity relation. $f_*/Z$
%$[Fe/H]<-1.5$, lowest Ursa Minor $[Fe/H]=-2.2$.
% \end{itemize}

\subsection{Missing Galactic satellites ?}

A powerful method to constrain models on the formation of the first
galaxies is to count the number density of simulated relic primordial
galaxies at z=0 and compare it with observations. Unfortunately there
are two main difficulties in doing this exercise correctly. (i)
Simulations of the formation of the first galaxies that include
realistic physics are computationally expensive and require a small
box size to achieve the mass resolution needed. This implies that we
can not evolve the simulation down to z=0. Therefore in order to
compare the simulation results (in this case at z=9) with observations
we need to extrapolate the results from z=9 to z=0 using other methods
(analytical or numerical). After reionization, \p3 galaxies should
stop forming and the ones already formed are expected to lose their
diffuse interstellar medium (ISM) due to photoevaporation. This should
stop their star formation.  Depending on the environment (cosmic
overdensity), a fraction of \p3 galaxies will merge to form larger
objects. Some larger objects will be stripped of part of their mass
due to tidal forces or shocks.  Therefore some dSphs could be \nor
galaxies observed during the phase of their disruption. (ii)
Observations could be missing a large fraction of the low-surface
luminosity density dwarfs. Another caveat that should be mentioned is
on the determination of the virial masses of dSphs, that could be
biased because of tidal interactions.

\begin{figure}[htb]
%\epsscale{0.8}
\plotone{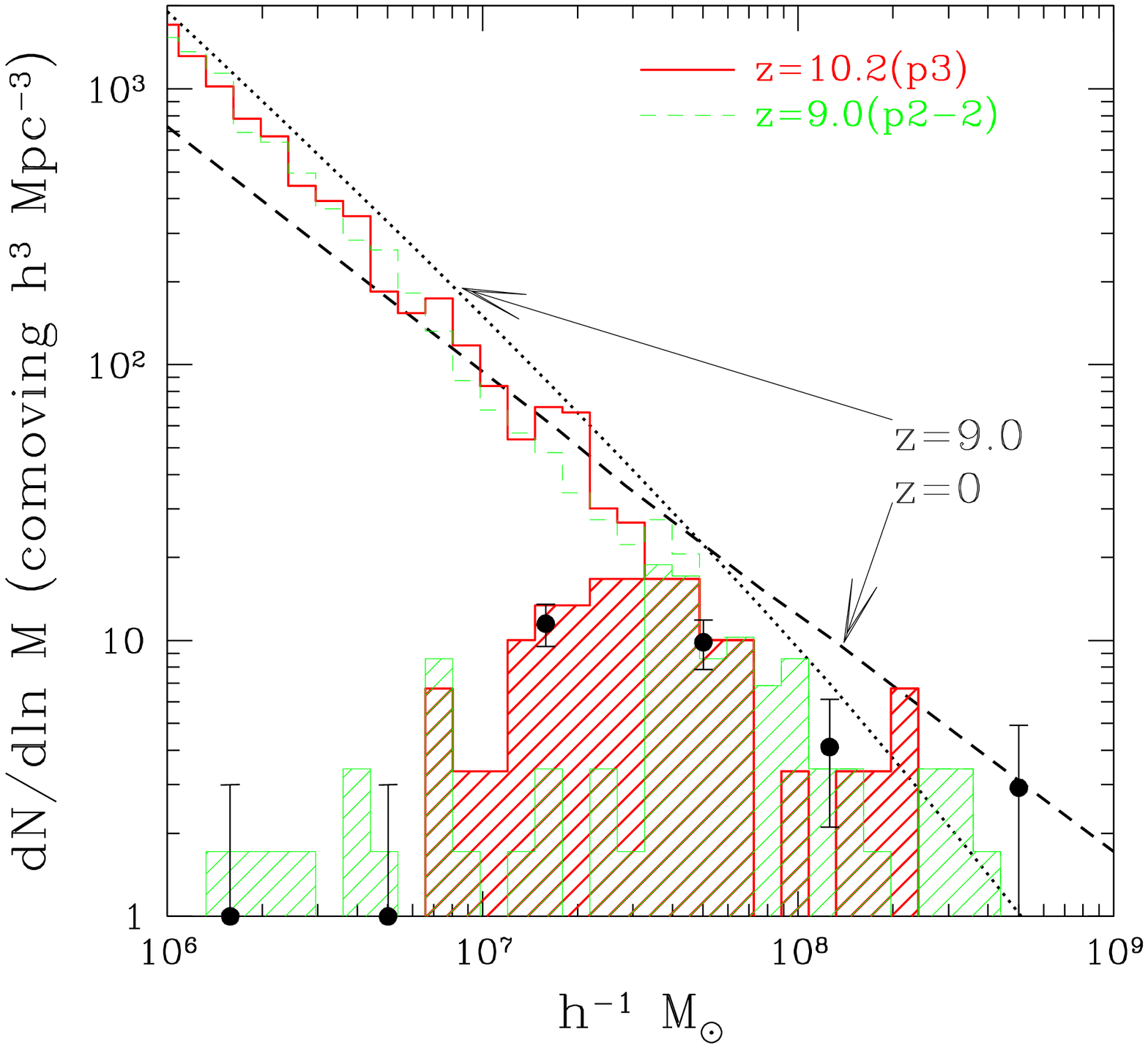}
\caption{\label{fig4} Mass function histogram of DM halos at $z \approx
  10$ for two simulations. The shaded histograms show the mass
  function of luminous galaxies (with $M_* \ge 10^4$
  M$_\odot$). The points with error-bars show the observed
  number density of galactic satellites at z=0. The thick dotted and dashed
  lines are shown to illustrate the evolution of the mass function
  from z=10 to z=0 according to the Press-Schechter formalism.}
\end{figure}

Fig.~\ref{fig4} shows the mass function of simulated galaxies at
about $z=9$ (shaded histograms) compared with observations at z=0 in
the Local Group. I show the result of two high resolution simulations
with \fesc$=0.01$, labeled ``p2-2'' and \fesc$=0.1$, labeled ``p3''
(see Table~\ref{tab:2}).

\section{Conclusions and Future Directions}

I have shown preliminary results, part of a larger work currently in
progress (Ricotti, Gnedin and Shull 2003, Ricotti and Gnedin 2003, in
preparation), to constrain and understand the theory for the formation
of the first galaxies in the universe. I have shown how observations
of dwarf galaxies in the Local Group can be used to constrain
theoretical results of cosmological simulations with radiative
feedback. If we establish a connection between dSphs and relics of the
first galaxies we can hope to learn in some detail the physics that
regulates the formation of the first stars in the universe, and the
importance of Population~III stars. The answer to currently popular
quests on the stellar initial mass function, stellar nucleosynthesis,
star formation efficiency of Population~III stars at $z \sim 30$,
could be found studying the most numerous an closest galaxies to the
Milky Way.

Studies of thermal and chemical evolution of the intergalactic medium
as a function of redshift and overdensity can further constrain the
model once a realistic treatment of supernova feedback is included in
the simulation (Ricotti, Gnedin and Shull 2003, in preparation).

% Deluxetable will work with this style file
% Plotone, plottwo, and plotfiddle should all work for plotting
%enter bibliography here
%.......................................................................
%\clearpage
\bibliographystyle{/home/ricotti/Latex/TeX/apj}
\bibliography{/home/ricotti/Latex/TeX/archive}

\end{document}